\documentclass[prl,preprint,groupedaddress,showpacs]{revtex4-1}

\usepackage{amsmath}
\usepackage{amssymb}
\usepackage{amsfonts}
\usepackage{mathrsfs}
\usepackage{epsfig}
\usepackage{bm}

\begin{document}

\title{Hosing Instability Suppression in Self-modulated Plasma Wakefields}
\author{J. Vieira$^{1,3}$}
\email{jorge.vieira@ist.utl.pt}
\author{W.B. Mori$^2$}
\author{P. Muggli$^3$}

\affiliation{$^1$GoLP/Instituto de Plasmas e Fus\~{a}o Nuclear-Laborat\'orio Associado,  Instituto Superior T\'{e}cnico, Universidade de Lisboa, Lisboa, Portugal}
\affiliation{$^2$Department of Physics and Astronomy, University of California Los Angeles, CA 90095, USA}
\affiliation{$^3$Max Planck Institute for Physics, Munich, Germany}
\today


\begin{abstract}
We show that the hosing instability can be suppressed after the saturation of the self-modulation instability of a long particle bunch if the plasma density perturbation is linear. 
We derive scalings for maximum bunch tilts and seeds for the self-modulation instability 
to ensure stable propagation beyond saturation of self-modulation. Numerical solutions of the reduced hosing equations and three-dimensional particle-in-cell simulations confirm our analytical findings. 
Our results may also apply when a train of particle bunches or laser pulses excite a linear wake. 
\end{abstract}

\maketitle

Plasma based acceleration is witnessing impressive advances~\cite{bib:patel_nature_2007}. Typical experiments use tightly focused ($\sigma_r\simeq 10~\mu\mathrm{m}$), high energy (1-100 J), ultra-short ($\sigma_z\simeq 10~\mu\mathrm{m}$) laser pulse (LWFA)~\cite{bib:tajima_prl_1979} or particle bunch (PWFA)~\cite{bib:chen_prl_1985}) drivers to excite relativistic, large amplitude plasma wakefields. These wakefields can accelerate electrons to high energies ($1-40~\mathrm{GeV}$) in short distances ($1-80~\mathrm{cm}$)~\cite{bib:wang_nat_2013,bib:blumenfeld_nat_2007}. To further increase energy gains, it was proposed to use existing high energy ($\gtrsim~10~\mathrm{kJ}$) short proton bunches ($\sigma_z\simeq 100~\mu\mathrm{m}$) as drivers to reach 600~GeV electron acceleration in 600~m long plasmas in the non-linear suck-in regime~\cite{bib:blue_prl_2003}. This scheme is called proton driven plasma wakefield accelerator (PDPWFA)~\cite{bib:caldwell_np_2009}.
 
The length of proton bunches currently available is $\sigma_z\simeq10~\mathrm{cm}$, much longer than the plasma wavelength ($\lambda_p$) even at low plasma densities ($n_0\simeq10^{14}-10^{16}~\mathrm{cm}^{-3}$). These long bunches are thus ideally suited to drive large acceleration gradients ($\sim1~\mathrm{GeV/m}$) through the self-modulation instability (SMI)~\cite{bib:kumar_prl_2010,bib:schroeder_prl_2011}, provided that  plasma ion motion is avoided~\cite{bib:vieira_prl_2012}. Unlike current PWFA experiments, which excite strongly non-linear wakes~\cite{bib:pukhov_apb_2002}, near-future PDPWFA experiments will operate in the linear PWFA regime. SMI experiments of long electron and positron bunches were also proposed to test key physics of the PDPWFA~\cite{bib:vieira_pop_2012}.

Hosing instability (HI), which can lead to bunch breakup~\cite{bib:lee_pof_1978,bib:whittum_prl_1991}, is considered a major impediment for the performance of the self-modulated (SM) PWFA. Previous work showed that during the linear stage of the SMI, the growth rate for the HI is similar to that of the SMI~\cite{bib:schroeder_pre_2012}. Bunch breakup could thus occur before SMI saturation. Moreover, the HI could lead to bunch breakup even after saturation of the SMI, where acceleration of externally injected particles can occur. The mitigation of the HI is therefore crucial for SM-PWFAs.

In this \emph{Letter} we show that the HI can be stabilized if the SMI can reach a fully saturated state and the density perturbations %
sustaining the wake are much smaller than the background plasma density, i.e. the wakefields are in the linear regime.
We also determine the amount of seeding required to reach SMI saturation before beam breakup occurs due to HI. We show that if this occurs 
stable wakefields in the SM-PWFA regime can be excited and maintained over long acceleration distances. We show that the suppression of  HI is similar to BNS damping of conventional accelerators~\cite{bib:bns}. Beam breakup due to HI can still occur when the density perturbations 
reach %
100\% in the blowout regime and in this case stabilization might be achieved through the use of a correlated energy spread. %
Numerical solutions of the reduced set of differential equations for the bunch centroid evolution and three-dimensional (3D) particle-in-cell (PIC) simulations with OSIRIS~\cite{bib:fonseca_book} confirm analytical findings.

We start by describing centroid displacements ($x_c$) of bunches with density profiles given by $n_b=(n_{b0} r_{b0}^2/r_b^2 ) \left[\Theta\left(r_b-r\right)+\delta\left(r-r_b\right)x_{c}\cos{\theta}\right]$, where $n_{b0}$ is the initial bunch density, $\Theta(x)$ is the Heaviside function, and where $\delta(x)$ is the Dirac delta function. The bunch radius is given by $r_b$, the transverse coordinates are $(x,y)$, $r_{b0}=r_b(z=0,\xi)$, $r=\sqrt{x^2+y^2}$ and the azimuthal angle is $\theta$. Hosing can be described by an integro-differential equation given in Ref.~\cite{bib:schroeder_pre_2012} which reduces to the following in the narrow bunch limit ($k_p r_b \ll 1$):
\begin{equation}
\label{eq:hosing}
\frac{\partial^2 x_c}{\partial z^2} = k_{\beta}^2\int_{-\infty}^{\xi} n_{\|}(\xi^{\prime}) k_p  \left[x_{\mathrm{c}}(\xi^{\prime}) - x_c(\xi) \right] \sin\left[k_p\left(\xi-\xi^{\prime}\right)\right]\mathrm{d}\xi^{\prime},
\end{equation}
where $k_{\beta}^2= k_p^2 m_e n_{b0}/(2 \gamma m_b n_0)$ and $k_p = \sqrt{4\pi n_0 e^2 / m_e}$ are the betatron and plasma wavenumbers, $e$ and $m_e$ are the electron charge and mass, and $q_b$, $m_b$, $\gamma=1/\sqrt{1-v_b^2/c^2}$ are the bunch particle's charge, mass, relativistic factor and $v_b$ its velocity. The co-moving frame coordinate, $\xi=z-v_b t$ 
is also used, where 
$\xi$ the location within the beam. The normalized longitudinal bunch density is $n_{\|}=r_{b0}^2/r_{b}^2$.

Similarly to the HI of laser pulses~\cite{bib:duda_pre_2000} it is possible to recast Eq.~(\ref{eq:hosing}) as:
\begin{subequations}
\label{eq:coupled}
\begin{align}
\left(\frac{\partial^2}{\partial z^2}+ \frac{e k_{\beta}^2}{q} \frac{\delta n_p}{n_0}\right) x_c & = k_{\beta}^2 x_w \\
\left(\frac{\partial^2}{\partial \xi^2}+ k_p^2\right) x_w & = k_p^2 n_{\|} x_c \label{eq:wakexc} \\
\left(\frac{\partial^2}{\partial \xi^2} + k_p^2\right) \frac{\delta n_p}{n_0} & = \frac{q_b k_p^2}{e} n_{\|} \label{eq:density},
\end{align}
\end{subequations} 
where we identify $\delta n_p/n_0$ as the plasma density perturbation and $x_w$ as the wake centroid.

Equations~(\ref{eq:coupled})a-c show that hosing occurs due to the coupling of $x_c$, which oscillates in $z$ at $k_{\beta}\left(\delta n_p/n_0\right)^{1/2}$, and the wake centroid, that oscillates in $\xi$ at $k_p$. The coupling term also includes the density of the beam. In addition, Eq.~(\ref{eq:density}) shows that $\delta n_p/n_0$ is described by a harmonic oscillator equation driven by the beam density. These equations are nonlinear. When linearized, i.e., when the beam density and the wake amplitudes are fixed, they provide various regimes of growth for hosing instability that can be identified by which of the first two equations in Eqs.~(\ref{eq:coupled}) are nearly resonant. When $x_w$ oscillates near $k_p$ then the instability is in the long pulse or strongly coupled regime and the beam centroid oscillates more rapidly than at its natural frequency. This is the regime of interest for current long beam research where the number of exponentiations scales as $\Gamma z \propto (n_{\|} k_{\beta}^2 z^2 k_p \xi)^{1/3}$.

\begin{figure}
\centering\includegraphics[width=\columnwidth]{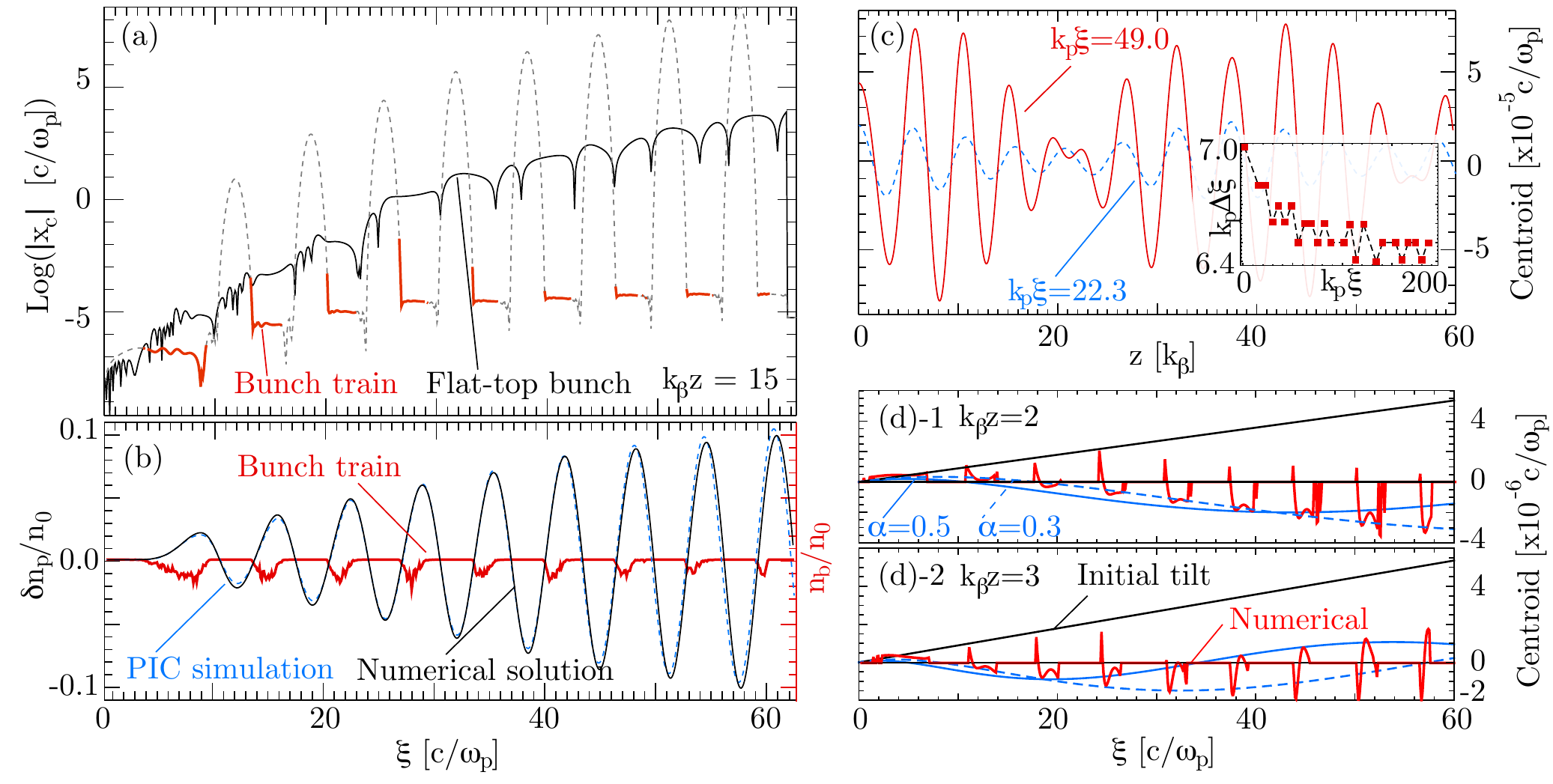}
\caption{(a) Numerical solutions for the centroid evolution for a flat-top bunch (black) and at the locations of a train of bunches (red). The solid gray line shows the centroid solution for the train of bunches at any $\xi$. (b) Corresponding bunch density profile (red), plasma density perturbations retrieved from a 3D PIC simulation (blue) and corresponding numerical solution (black) using Eq.~(\ref{eq:density}). (c) Numerical solution for $x_c$ for a train of Dirac-delta like bunches (with length $k_p \sigma_z=0.1$ and density $n_l = n_0/(k_p \sigma_z)$) located at maximum focusing field regions. The inset shows the relative distance between beamlets. (d)1-2 Numerical (red) and analytical (blue) solutions for the centroid evolution of a collection of beamlets. The initial centroid displacement is $x_{c0} = 8.93\times10^{-8}\xi$.}
\label{fig:numerical-analytical}
\end{figure}

If one wants to couple SMI with HI an additional equation that relates the evolution of the beam density to the wake amplitude (plasma density perturbation) is needed~\cite{bib:schroeder_pre_2012}. However, the coupling of a fully self-modulated beam or a train of beamlets with HI can be studied without this additional equation. Analysis of Eq.~(\ref{eq:coupled}) indicates that a train of bunches can propagate without significant centroid oscillations during long propagation distances ($k_{\beta} z\gg 1$). This is demonstrated in Fig.~\ref{fig:numerical-analytical}a which compares numerical solutions to Eq.~(\ref{eq:hosing}) or Eq.~(\ref{eq:coupled}) for an electron bunch with a non-evolving flat top density profile (with $n_{\|}=1$) with a sharp rise  (black) and for a train of bunches (red) after $k_{\beta} z = 15$. The bunch train profile is shown in Fig.~\ref{fig:numerical-analytical}b (red) and it was taken from a fully self-consistent 3D PIC simulation (to be described in more detail later) after a long beam had undergone the SMI. For the bunch train (Fig.~\ref{fig:numerical-analytical}a) the centroid displacements are up to 6-7 orders of magnitude smaller than for the flat density case. This demonstrates the possible suppression of hosing for a fully self-modulated beam.

Hosing of a fully self-modulated beam or a train of bunches cannot be described in terms of the often quoted asymptotic solutions and regimes such as the long pulse/strongly coupled regime. As SMI occurs and $n_{\|}$ becomes modulated the coupling term on the right hand side of Eq.~(\ref{eq:wakexc}) leads to harmonic generation of $x_c$ due to its nonlinearity. This leads to harmonic generation of each quantity. As the number of harmonics increases the analysis of the interaction between HI and SMI becomes %
more %
difficult to analyze. 

Therefore,  to understand the main physical mechanisms leading to hosing suppression in the wake driven by a train of bunches we consider a simplified particle model where $n_{\|}$ is given by $n_{\|}^{\mathrm{SM}}=\sum_{l=0}^m k_p^{-1} n_l \delta\left(\xi -\xi_l\right)$, where $\xi_{l}$ is the location in $\xi$ of the $l^{\mathrm{th}}$ beamlet (or particle), and where $k_p^{-1} n_l$ is proportional to its charge. Similar models were employed to investigate beam-break up instabilities in RF linear accelerators~\cite{bib:particlemodel} and to study trapped particle instabilities in plasma waves~\cite{bib:kruer_prl_1969}. Inserting $n_{\|}^{\mathrm{SM}}$ into Eq.~(\ref{eq:coupled}) yields: 
\begin{equation}
\label{eq:hi-smb}
\frac{\partial^2 x_m }{ \partial z^2} + x_m(z) k_{\beta}^2 \left(\frac{\delta n_p}{n_0} \frac {q}{e}\right)_m = k_{\beta}^2 \sum_{l=0}^{m} n_l w_l x_l(z),
\end{equation}
where $x_m=x_c(z,\xi_m)$, $\left(\delta n_p /n_0\right)_m = \sum_{l=0}^{m} n_l \sin\left[ k_p\left( \xi_l-\xi_m \right)\right]$ is the amplitude of the plasma density fluctuations at $\xi = \xi_m$, and $w_l=\sin\left[ k_p\left( \xi_l-\xi_m \right)\right]$ is a weighting factor. Equation~(\ref{eq:hi-smb}) shows that the centroid of each beam is described by a driven harmonic oscillator equation and that each centroid oscillates at a frequency, $k_{m}^2=k_{\beta}^2\left(\frac{q}{e} \delta n /n_0\right)_m$. The driving term is the weighted centroid oscillations of the preceding beamlets, $\sum_{l} n_l w_l x_l$.

In order to understand how hosing of a train of short bunches evolves, we examine the first few terms in Eq.~(\ref{eq:hi-smb}) with $n_l = 1$, i.e. implicitly assuming that the wake perturbation is enough to guide the beam with given emittance. For the first beamlet $\partial^2 x_{0}/\partial z^2 = 0$, and if $\partial x_0/\partial z = 0$ at $z=0$, then $x_0=x_{c0}$. For the second beamlet $\partial^2 x_1/\partial z^2 + k_{\beta}^2 \frac{q}{e} \sin{[k_p(\xi_0-\xi_1])} (x_1-x_0)=0$. For $x_0 = 0$ and $\partial x_1 / \partial z= 0$ then $x_1 = x_{1c} \cos\left[z k_{\beta}\sqrt{\frac{q}{e}\sin{[k_p(\xi_0-\xi_1)]}}\right]$. Bounded  $x_1$ centroid oscillations occur when the second beamlet resides in focusing field regions, i.e. when $\frac{q}{e}\sin{[k_p(\xi_0-\xi_1)]}>0$, for which the beamlet oscillates at $k_1 = k_{\beta}\sqrt{\frac{q}{e}\sin{[k_p(\xi_0-\xi_1)]}}$. The equation of motion for the third beam centroid is therefore $\partial^2 x_2 / \partial z^2 + k_{\beta}^2 x_2 \frac{q}{e} (\sin{[k_p(\xi_0-\xi_2)]}+\sin{[k_p(\xi_1-\xi_2)]}) = w_1 k_{\beta}^2 x_{1c} \cos\left[z k_{\beta}\sqrt{\frac{q}{e}\sin{[k_p(\xi_0-\xi_1)]}}\right] $ where $w_1 = \sin{[k_p(\xi_1-\xi_2)]}$. Bounded oscillation for $x_2$ also requires that the third beamlet resides in focusing field regions such that $\frac{q}{e}(\sin{[k_p(\xi_0-\xi_2)]}+\sin{[k_p(\xi_1-\xi_2)]})>0$. In this case, the equation of motion  for $x_3$ is a  driven harmonic oscillator. The driving term oscillates at $k_1 = k_{\beta}\sqrt{\frac{q}{e}\sin{[k_p(\xi_0-\xi_1)]}}$. The natural frequency for the centroid oscillation is given by $k_2^2 = k_{\beta}^2 (\frac{q}{e}\sin{[k_p(\xi_0-\xi_2)]}+\sin{[k_p(\xi_1-\xi_2)]}$. Hence, ensuring that $k_2 \ne k_1$ avoids resonant $x_2$ oscillations and rapid growth. By extending this argument to the following beamlets it can be recognized that avoiding resonant centroid growth requires that every beamlet oscillate at a different frequency. This condition is naturally satisfied for fully self-modulated bunches in the linear regime, where each beamlet resides in focusing field regions and the amplitude of the focusing field (and the wavenumber) increases for each beamlet. The natural frequency can also vary if the spacing between bunches varies or if there is an energy chirp on the beam ($k_{\beta}\propto1/\sqrt{\gamma}$).

To illustrate the suppression of hosing, we present in Fig.~\ref{fig:numerical-analytical}c numerical solutions for a case where short (nearly delta function) beamlets are in regions of maximum focusing field, as would be expected in a self-modulated scenario. We use Eq.~(\ref{eq:coupled}) for very short bunches, which is then equivalent to Eq.~(\ref{eq:hi-smb}). Figure~\ref{fig:numerical-analytical}c demonstrates that the amplitude of $x_c$ remains close to $x_{c0}$ for $k_{\beta} z\gg 1$. These conclusions are consistent with the results for the more realistic bunch train density distributions of Figs.~\ref{fig:numerical-analytical}a-b. Figure~\ref{fig:numerical-analytical}b, which superimposes the position of each beamlet in the wake it excites, then shows that the bunch train considered in Fig.~\ref{fig:numerical-analytical}a is in regions of focusing fields. More details are in the figure caption.

Numerical solutions of Eq.~(\ref{eq:coupled}) show that finding separations $\Delta \xi_m$ between beamlets that ensure they reside in maximum focusing fields is challenging as $\Delta \xi_m$ depends on their relative position within the bunch train, on their length, charge, and density profile. Therefore, producing a train of bunches with the conditions to suppress the HI while making a wake over large distances would be very challenging experimentally. For example, the inset of Fig.~\ref{fig:numerical-analytical}c shows the optimal spacing for beamlets for one case. The spacing varies and each is longer than $\lambda_p$. However, long bunches can self-consistently evolve into this optimal configuration through SMI because 
the process of self-modulation itself requires the bunches reside in focusing regions. This can be seen in Fig.~\ref{fig:numerical-analytical}b which also shows that in this case linear wakes are still excited during the non-linear stage of SMI. To demonstrate this, we show in Fig.~\ref{fig:numerical-analytical}b, that the numerical solution to Eq.~(\ref{eq:density}) using the simulation $n_{\|}$ and the corresponding $\delta n_p/n_0$ retrieved directly from the simulation are nearly indistinguishable.

It is also possible to obtain analytical expressions for the evolution of $x_m$ in self-modulated regimes if we simplify Eq.~(\ref{eq:hi-smb}) by assuming a constant $w_l=\alpha$ and $k_m^2 = \alpha k_{\beta}^2 \sum_l n_l$. This approximation corresponds to wakefields growing secularly along the bunch and to each beamlet equally driving the oscillations of following beamlets. The value of $\alpha$ depends on the exact density profile of each beamlet. Purely analytical predictions for the detailed bunch profile during the non-linear stage of the SMI, however, are currently unavailable. Under the above assumptions we can replace the sums by integrals whereby Eq.~(\ref{eq:hi-smb}) becomes:
\begin{equation}
\label{eq:hosing_modulated}
\left(\frac{\partial^2}{\partial z^2} + \alpha k_{\beta}^2 \int_{\infty}^{\bar{\xi}} k_p n_{\|}(\xi^{\prime}) \mathrm{d}\xi^{\prime}\right)x_c = \alpha k_{\beta}^2 \int_{-\infty}^{\bar{\xi}} k_p x_c(\xi^{\prime}) n_{\|}(\xi^{\prime}) \mathrm{d}\xi^{\prime},
\end{equation}
where $\bar{\xi}= \xi \sigma_z/\Delta \xi$ is a new variable that runs only through the regions where $n_{\|}\ne 0$. We can solve Eq.~(\ref{eq:hosing_modulated}) for beamlets with $n_{\|}=1$ (i.e. assuming that the charge on each beamlet is constant) by differentiating Eq.~(\ref{eq:hosing_modulated}) once in $\bar{\xi}$ and by solving the resulting equation for $\partial x_c /\partial \bar{\xi}$ using $x_{c0}=\delta_{\mathrm{HI}} \bar{\xi}$ and $\partial x_{c0}/\partial z = 0$ yielding: 
\begin{equation}
\label{eq:hosing_flat}
k_p x_c =\frac{2  \delta_{\mathrm{HI}} }{\alpha k_{\beta}^2 z^2}\left[-1+\cos\left(N_{\mathrm{flat}}\right)+ N_{\mathrm{flat}}\sin\left(N_{\mathrm{flat}}\right)\right],
\end{equation}
where $N_{\mathrm{flat}} = k_{\beta} z \sqrt{\alpha k_p \bar{\xi}}$. Equation~(\ref{eq:hosing_flat}) indicates that $k_p x_c\propto\delta_{\mathrm{HI}}\sqrt{\alpha k_p \bar{\xi}}/(k_{\beta} z)$ demonstrating HI damping after SMI saturation for $k_{\beta} z\gg 1$. 
In Fig.~\ref{fig:numerical-analytical}d(1-2) we compare numerical solutions of Eq.~(\ref{eq:coupled}) using the $n_{\|}$ distribution shown in Fig.~\ref{fig:numerical-analytical}b with  Eq.~(\ref{eq:hosing_flat}) for $w_l=0.3$ or $w_l=0.5$ and with constant $n_{\|}$ that is the average of the actual distribution. Figure~\ref{fig:numerical-analytical}d shows that the numerical solution for $x_c$ varies within each beamlet because the betatron frequency is $\xi$-dependent within each bunch. The derivation of Eq.~(\ref{eq:hosing_flat}) assumes that the betatron frequency is constant (i.e., at a value of $k_{\beta} \alpha^{1/2}$) and hence does not take this effect into account. Nevertheless there is agreement with the peaks of the numerical solution for $x_c$ and the analytical solution for $k_{\beta} z\lesssim 3$.  For $k_{\beta} z\gtrsim 4$ the agreement is worse because the assumption of a constant $\alpha$ becomes progressively worse. We also note that the values for the weights $w_l$ vary between $0.8$ and $0.2$ from the front to the back of the bunch train.

Our analysis shows that hosing suppression in self-modulated regimes occurs because the betatron frequency of each bunch varies along the train. The secular growth of the plasma focusing force responsible for HI suppression is intrinsic to plasma wakefields in the linear regime. In the non-linear wakefield regime driven by negatively charged bunches, the focusing fields are set by the ion column density. Thus, all beamlets oscillate at the same frequency, and hosing can still grow. Simulations demonstrate that electron bunches with a correlated 
energy spread could nevertheless be used to damp/suppress hosing in this case %
because $k_{\beta}\propto1/\sqrt{\gamma}$ now varies along the bunch. %
In addition, when the drive bunches are not short when compared with $\lambda_p$, as is often the case in practice, the variation of the accelerating field across each bunch could also naturally contribute to damp HI. For positively charged bunches, most of the bunch defocuses during the growth of SMI in the blowout regime. Thus, ensuring stable propagation for positively charged drivers requires wake excitation in the linear regime. The HI suppression mechanisms mentioned above are similar to BNS damping in RF linear accelerators~\cite{bib:bns}.

We can estimate the peak bunch density after SMI saturation considering that $n_{b0} r_{b0}^2=n_{b(\mathrm{sat})} r_{b(\mathrm{sat})}^2$, where $n_{b/(\mathrm{sat})}$ and $r_{b(\mathrm{sat})}$ refer to the matched bunch density and radius after SMI saturation~\cite{bib:referee}. By assuming that the wake grows secularly along the fully self-modulated bunch, the matched bunch radius in the linear regime becomes $r_{b(\mathrm{sat})}\simeq[2\epsilon_N^2 / k_p^2 (n_{b0}/n_0)(m_e/m_b)(\lambda_p/\sigma_z)]^{1/4}$. Thus $n_{b(\mathrm{sat})}\simeq n_{b0} r_{b0}^2 [(\gamma k_p^2/2\epsilon_N^2) (n_{b0}/n_0) (m_e/m_b) (\sigma_z/\lambda_p)]^{1/2}$, where $\epsilon_N$ is the normalised emittance. For the parameters of future PDPWFA experiments with $n_{b0}/n_0\simeq 10^{-2}$, $\sigma_z \simeq 10~\mathrm{cm}$, $n_0\simeq 10^{14}~\mathrm{cm}^{-3}$, $r_{b0}\simeq200~\mu\mathrm{m}$ and $\epsilon_N\simeq3\mathrm{mm\cdot mrad}$ this gives $n_{b(\mathrm{sat})}\simeq 1.7>1$. Cylindrically symmetric PDPWFA simulations, however, show that emittance can increase by an order of magnitude. In that case, $n_{b(\mathrm{sat})}\simeq 0.17$. Thus, the wakefield will still be in the linear regime in future PDPWFA experiments, and hosing can be stabilised after SMI saturation.

The condition for stable wake excitation in SM-PWFA is that SMI saturation occurs well before bunch break-up due to HI. Since HI and SMI have similar growth rates, this condition can be fulfilled when the seed for SMI is larger than for HI, i.e., the initial focusing force that seeds hosing ($\langle W_{\perp,\mathrm{HI}} \rangle$) needs to be smaller than that seeding self-modulation ($\langle W_{\perp,\mathrm{SMI}} \rangle$). This is the same as having the seed for $x_w$ being smaller than the seed for $\delta n_p/n_0$. Among several SMI seeding mechanisms~\cite{bib:vieira_inpreparation,bib:mori_prl_1992,bib:schroeder_pop_2013} we consider seeding by bunches with short rise times~\cite{bib:fang_prl_2013} for which $\langle W_{\perp,\mathrm{SMI}} \rangle \propto k_{\beta}^2 k_p r_{b0}$. In addition, beam tilts that seed hosing lead to $\langle W_{\perp,\mathrm{HI}} \rangle =k_{\beta}^2 k_p \sigma_z \delta_{\mathrm{HI}}$. Hence the stable propagation condition $\langle W_{\perp,\mathrm{SMI}} \rangle /\langle W_{\perp,\mathrm{HI}} \rangle \gtrsim 1$ holds as long as $\delta_{\mathrm{HI}}\lesssim r_b/\sigma_z$ ($x_{c0}=\delta_{\mathrm{HI}}\xi$), consistent with~\cite{bib:schroeder_pop_2013}. Stable propagation then occurs for all beamlets whose centroid initially resides within the bunch radius, at the bunch front.

\begin{figure}
\centering\includegraphics[width=\columnwidth]{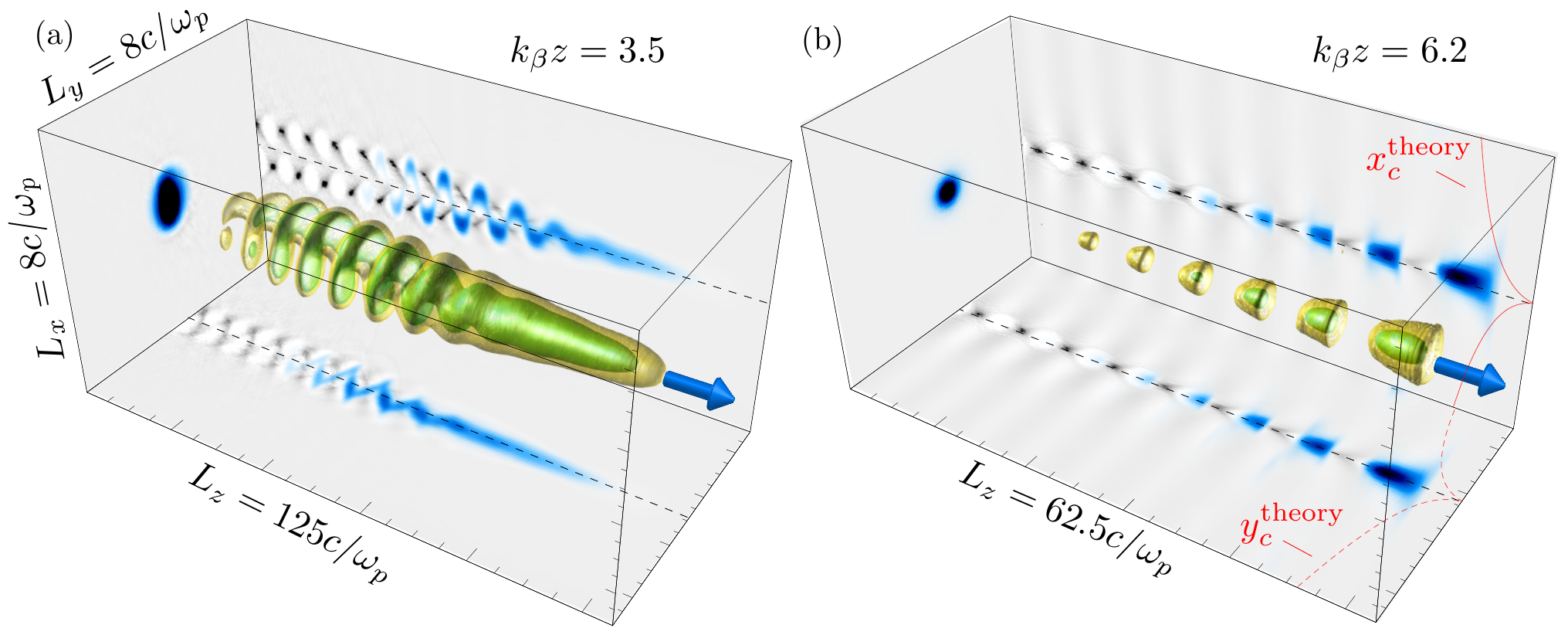}
\caption{OSIRIS simulation results: bunch density iso-surfaces (yellow and green). Projections show electron bunch (blue) and plasma density (gray). (a) Bunch with smooth temporal profile (b) Bunch with sharp rise long fall current profile for SMI seeding. The red lines on side projections show HI theoretical predictions for non self-modulated bunches.}
\label{fig:seed-hosing}
\end{figure}

A set of 3D PIC simulations were performed with %
the numerical code %
OSIRIS~\cite{bib:fonseca_book}. See Supplemental Material at [URL] for the simulation parameters. Figure~\ref{fig:seed-hosing}a illustrates the transition between the linear stage of hosing and to a nonlinear couplingng between the SMI and the HI. In this case, even with a very small seed for the HI ($k_p \delta_{\mathrm{HI}}=0.001$) and essentially no SMI seeding, HI strongly breaks up the bunch density after a short propagation distance ($k_{\beta}z=3.5$) and before SMI can grow. Figure~\ref{fig:seed-hosing}b shows results from the half bunch.  Even with an initial HI seed ten times larger than in the case of Fig.~\ref{fig:seed-hosing}a ($k_p \delta_{\mathrm{HI}}=0.01$), the bunch is free of HI. The beam becomes fully self-modulated and then stably propagates over a longer distance into the plasma ($k_{\beta}z=6.2$). In this case, existing hosing theory for flat bunches significantly overestimates $|x_c|$. 
We note that the beam charge in the modulated bunch in Fig.~\ref{fig:seed-hosing}b is four times smaller than in Fig.~\ref{fig:seed-hosing}a. This lower charge is because the bunch is cut on the middle, and half the charge is defocused by the SMI. The propagation distance in Fig.~\ref{fig:seed-hosing}b, however, is twice as large as for Fig~\ref{fig:seed-hosing}a. The theoretical number of e-foldings in Fig.~\ref{fig:seed-hosing}a should then be similar to that of Fig.~\ref{fig:seed-hosing}b, since the number of e-foldings scale as $(n_b k_p\xi k_{\beta}^2 z^2)^{1/3}$. However, centroid oscillations are much larger in Fig.~\ref{fig:seed-hosing}a than in Fig.~\ref{fig:seed-hosing}b. In addition, simulations show that whereas the half cut bunch continues to propagate stably in Fig.~\ref{fig:seed-hosing}b, beam breakup occurs in the case of Fig.~\ref{fig:seed-hosing}a. Together with the fact that the initial tilt that seeds the hosing in Fig.~\ref{fig:seed-hosing}b is ten times larger than that in ~\ref{fig:seed-hosing}a, Fig.~\ref{fig:seed-hosing} shows that hosing suppression can occur in the linear regime after SMI saturation.

\begin{figure}
\centering\includegraphics[width=\columnwidth]{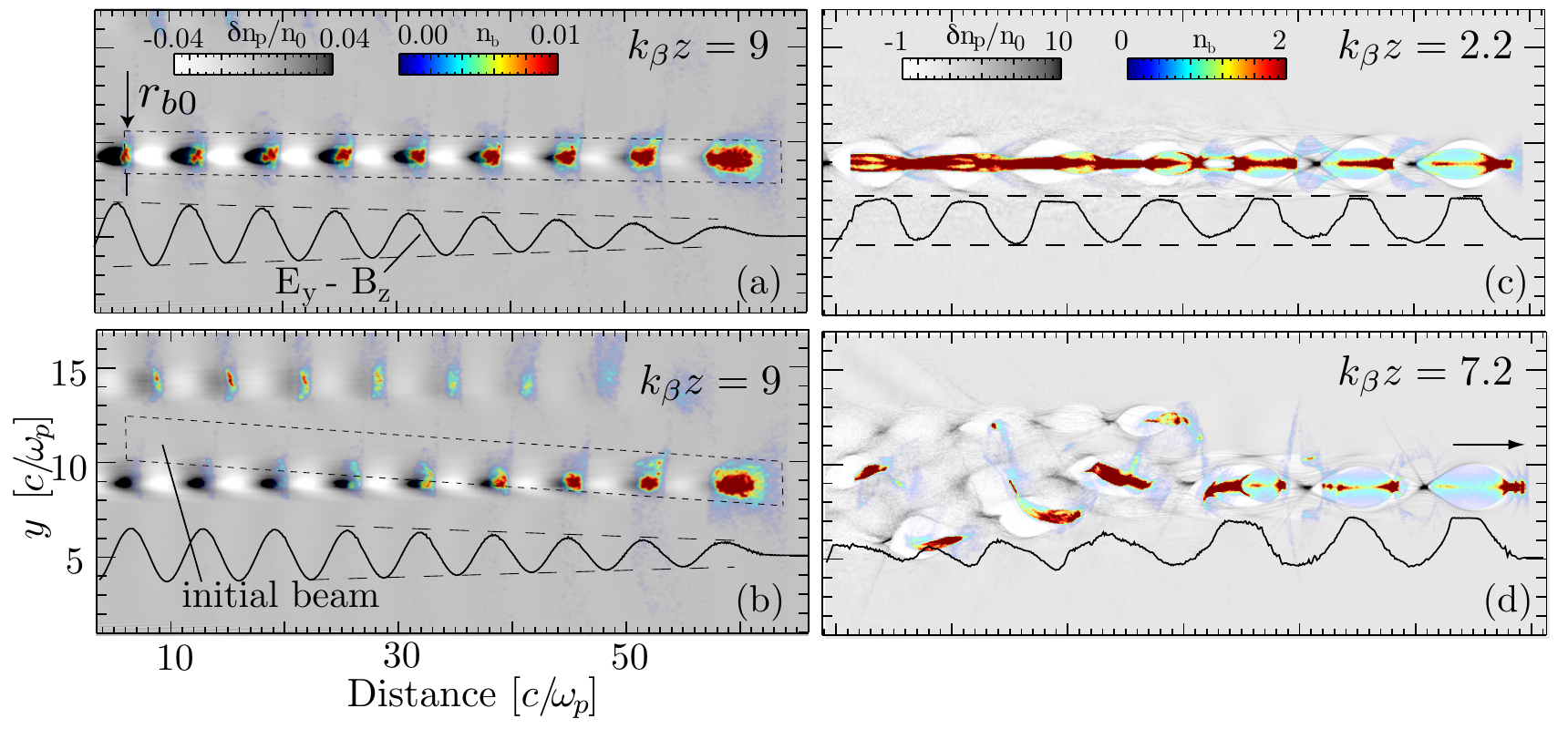}
\caption{Osiris simulation results of the propagation of the long electron bunch (blue-red colors) in a plasma (gray colors) in the linear ((a) and (b)) and non-linear  ((c) and (d)) wakefield regime. The shape of the initial bunch profile is shown by the short dashed lines. The initial bunch radius $r_{b0}$ is also indicated. Plasma focusing force $E_y-B_z$ (solid lines) and envelope of $E_y-B_z$ (long dashed-lines) are also illustrated.}
\label{fig:non-linear}
\end{figure}

Figures~\ref{fig:non-linear}a-b show that hosing suppression also occurs for flat top bunches with different initial tilts and with $n_b/n_0=0.01$ so long as the wake is still in the linear regime %
after saturation. %
Figure~\ref{fig:non-linear}a shows results from a simulation in which the tilt was small enough so that centroid variation across the entire beam was less than the initial spot size, i.e., where $x_{c0}=\delta_{\mathrm{HI}}\sigma_z =0.6 r_b < r_b$. In this case all the self-modulated beamlets propagate stably. The y direction focusing force ($W_y=E_y-B_z$) increases along the bunch, resulting in betatron frequency detuning among self-modulated beamlets, which leads to HI suppression and which is consistent with Eqs.~(\ref{eq:hi-smb}) and (\ref {eq:hosing_modulated}). The simulation results from Fig.~\ref{fig:numerical-analytical}b correspond to on-axis lineouts from Fig.~\ref{fig:non-linear}a. In Fig.~\ref{fig:non-linear}(b), results for a beam with a larger tilt are shown. In this case $x_{c0} = \delta_{\mathrm{HI}} \sigma_z = 3 r_b > r_b$ and only beamlets satisfying $\xi \lesssim r_b/\delta_{\mathrm{HI}}$ ($x_{c0}<r_b$) propagate stably (i.e. those with $k_p \xi \gtrsim 30$), %
in agreement with analytical scalings. %
Also in agreement with theory, additional simulations (not shown) also confirm these conclusions for positron bunches.

Figure~\ref{fig:non-linear}c-d illustrate the breakup of the same bunches as used in Figs~\ref{fig:non-linear}a-b, but with $n_b/n_0=0.5$ such that the wakefields driven by the SMI eventually reach the nonlinear blowout regime. As the bunch self-modulates, the amplitude of the plasma focusing force becomes constant throughout the entire bunch train and the amplitude is the same for each bunch (solid and dashed lines in Fig.~\ref{fig:non-linear}c). As discussed earlier this prevents the suppression of the HI and the beam is seen to eventually break apart  due to resonant HI growth (Fig.~\ref{fig:non-linear}d). Other wakefield saturation mechanisms (e.g. due to fine scale mixing of electron trajectories~\cite{bib:dawson_pr_1959}) could also lead to HI growth.

In conclusion, we have shown that the hosing instability of long particle beams can be suppressed and stabilized if the beam first becomes fully self-modulated and the resulting wake and density perturbations remain in the linear regime. This requires that the seed for SMI is larger than for HI. Fully self-consistent PIC simulations show that for long particle beams with sharp rise times the beam can propagate for long distances exciting a wakefield that could be used to accelerate externally injected particles. This suppression mechanism is analogous to BNS damping in conventional linear accelerators. These results should also apply to a train of laser pulses and this will be addressed in future work.


\acknowledgements
Work supported by FCT (Portugal), grant EXPL/FIS-PLA/0834/1012; by the European Research Council (ERC-2010-AdG Grant 267841); by DOE grants  DE-SC0008491, DE-SC0008316,and DE-FG02- 92-ER40727. J.V. acknowledges the Alexander von Humboldt Foundation for a post-doctoral fellowship at the Max Planck Institute for Physics in Munich. We acknowledge PRACE for access to resources on J\"{u}Queen (J\"{u}lich) and SuperMUC (Leibniz Research Center) and INCITE for use of the Titan supercomputer. We also acknowledge useful discussions with Prof. L. O. Silva and Prof. R. A. Fonseca.

\end{document}